# Crystal growth and metallic ferromagnetism induced by electron doping in FeSb$_2$

Mohamed A. Kassem[1,2], Yoshikazu Tabata[1], Takeshi Waki[1], Hiroyuki Nakamura[1]

[1] Department of Materials Science and Engineering, Kyoto University, Kyoto 606-8501, Japan,
[2] Department of Physics, Faculty of Science, Assiut University, Assiut 71516, Egypt,

**Abstract:**

In order to study the metallic ferromagnetism induced by electron doping in the narrow-gab semiconductor FeSb$_2$, single crystals of FeSb$_2$, Fe$_{1-x}$Co$_x$Sb$_2$ ($0 \leq x \leq 0.5$) and FeSb$_{2-y}$Te$_y$ ($0 \leq y \leq 0.4$), were grown by a simplified self-flux method. From powder x-ray diffraction (XRD) patterns, wavelength-dispersive x-ray spectroscopy (WDX) and x-ray Laue diffraction, pure and doped high-quality single crystals, within the selected solubility range, show only the orthorhombic *Pnnm* structure of FeSb$_2$ with a monotonic change in lattice parameters with increasing the doping level. In consistence with the model of nearly ferromagnetic small-gap semiconductor, the energy gap of FeSb$_2$ Pauli paramagnet gradually collapses by electron doping before it closes at about $x$ or $y \simeq 0.15$ and subsequent itinerant electron anisotropic ferromagnetic states are observed with higher doping levels. A magnetic phase diagram is established and discussed in view of proposed theoretical scenarios.

## 1. Introduction

Intermetallic compounds are usually metallic. However, some compounds including the rare-earth based typical Kondo insulators such as $SmB_6$ and $YbB_{12}$ intermetallics and the proposed transition-metal counterparts FeSi and $FeSb_2$ show a semiconductor behavior and are described as strongly correlated semiconductors[1–4]. In these compounds, generally, itinerant and localized electrons hybridize and open a narrow gap at the Fermi level with a large density of states[5]. As many iron- and ruthenium-based semiconductors, $FeSb_2$ has attracted a considerable attention because of its unusual transport and magnetic behavior as well as its notable thermoelectric properties[4,6,7]. $FeSb_2$ and FeSi were suggested as examples of non-rare-earth based Kondo semiconductors[8]. Moreover, according to the recent theoretical study, $FeSb_2$ is proposed as a nearly ferromagnetic small gap semiconductor and one can switch from a small gap semiconductor to a ferromagnetic metal with magnetic moment $\approx 1\mu_B$ per Fe ion by external magnetic field[9].

Common nonmagnetic insulating state with thermal activation-type magnetization process have been observed for poly- and single-crystalline samples of $FeSb_2$. However, some previously presented physical properties of $FeSb_2$ depend on the sample and has been elucidated as very sensitive to ppm-level impurities. For example, anisotropic insulating behavior with a metal-insulator transition at 50 K along the *b*-axis observed by Petrovic *et al.* contrasts the almost isotropic insulating behavior down to 10 K along the three principal axes observed by other groups[4,7,10]. Further, a reported colossal Seebeck coefficient of – 45 mV K$^{-1}$ at 10 K observed by Bentien *et al.* has not been observed in single crystals prepared by others[6,11]. The questionable reproducibility in the large Seebeck effect has recently attributed to a phonon-drag effect[7].

The magnetization *M* of FeSi (cubic B20 structure) and $FeSb_2$ (orthorhombic marcasite structure) shows an activation-type increase with increasing temperature *T* above 70 and 50 K, respectively[12,13]. After showing a broad maximum around 500 and 300 K, respectively, the *M-T* curve exhibits a Curie-Weiss like behavior[12,13]. In order to explain the magnetization process of FeSi, a model was proposed in Takahashi's spin fluctuation theory that employed a simple density of states curve with an energy gap $E_g$ and considered quantum spin fluctuations together with thermal spin fluctuations[14,15]. The model has succeeded to evaluate $E_g$ (~ 62 and 30 meV for FeSi and $FeSb_2$, respectively) from the experimental magnetization data and to predict its pressure-dependence in qualitative consistence with results of resistivity measurements[16–18].

On the other hand, the Kondo description of $FeSb_2$ suggests that the substitution of third element can give rise to drastic changes in the electronic and magnetic states[19]. Emergence of ferromagnetic and antiferromagnetic states have been observed by substitutions with Co and Cr for Fe and by Te for Sb[20–22]. However, as well as the properties of $FeSb_2$, the reproducibility of its solid solutions properties needs investigation[4,20–22]. Further, the study of the energy gap evolution by doping, estimated from magnetization based on the itinerant-electron magnetism scenario proposed in Takahashi's spin fluctuation theory, would imply how much the model fits to explain the experimental results and hence the origin of magnetization process in this system.

Here, we report simple self-flux crystal growth of high-quality $FeSb_2$ and its electron-doped solid solutions, $Fe_{1-x}Co_xSb_2$ ($0 \leq x \leq 0.5$) and $FeSb_{2-y}Te_y$ ($0 \leq y \leq 0.4$). The evolution of the energy gap to metallic ferromagnetic states by electron doping is studied from a viewpoint of Takahashi's theory of itinerant electron magnetism. A magnetic phase diagram of electron doped $FeSb_2$ is presented.

## 2. Materials and Methods

Large single crystals of $FeSb_2$, $Fe_{1-x}Co_xSb_2$ ($0 \leq x \leq 0.5$) and $FeSb_{2-y}Te_y$ ($0 \leq y \leq 0.4$) were synthesized by a high temperature Sb self-flux method. Lumps of Fe (99.95 % Alfa Aesar) and Co (99.9 % Kojundo Chemical Laboratory Co., ltd.) and grains of Sb (99.999 % Kojundo Chemical Laboratory Co., ltd.) and Te (99.999 % Kojundo Chemical Laboratory Co., ltd.) have been used. A mixture of Fe: Sb, $(Fe_{1-x}Co_x)$: Sb or Fe: $(Sb_{2-y}Te_y)$ with a molar ratio equal to 8: 92 has been sealed without crucibles in an evacuated quartz tube (1 cm in diameter) with a dimple. The ampule was heated in a muffle furnace and kept at 1000°C for 2 hours then cooled down to 800°C over 10 hours and finally slowly cooled down to 650°C over 100 hours. The ampule has been taken out from the furnace at 650°C and subsequently turned upside down to separate the flux liquid from grown crystals. The excess flux attached to the surface of picked-up crystals has been removed by a further sealing in a long quartz ampule heated for 24 hours in a temperature-gradient horizontal furnace with the sample at the highest temperature of 800°C.

Crushed powder of a part of the grown crystals were investigated at room temperature by powder XRD (PANalytical, X'Pert Pro) with Cu $K_{\alpha1}$ radiation monochromated by a Ge (111)-Johansson-type monochromator. The crystal structure parameters were refined by the Rietveld refinement analysis using TOPAS (Bruker) software. The actual chemical compositions were investigated by SEM-WDX (Hitachi, S-3500H). The crystallinity level and crystalline axes were determined by the x-ray Laue method.

Magnetization and magnetic anisotropy were measured in a temperature range of $T$ = 2 - 350 K using a SQUID magnetometer (Quantum Design, MPMS-7) installed in the LTM center of Kyoto University. Three single crystals of $FeSb_2$ were cut and polished in rectangular rod shapes along the different principal crystal-axes for electrical resistivity, $\rho(T)$, measurements. $\rho(T)$ was measured at zero magnetic field in a temperature range of $T$ = 4 - 300 K by employing the four-probe method.

## 3. Results and discussion

Grown single crystals with the long axis parallel to the $b$ axis and others with large {110} faces (as shown in the inset of Fig. 1(a)) are large enough for different macroscopic investigations. The synthesized phase and crystal structure of the grown crystals were investigated by powder XRD for crushed parts. The observed XRD patterns shown in Fig. 1(a) are fitted by the Rietveld method using the structure parameters of $FeSb_2$ marcasite structure of *Pnnm* orthorhombic symmetry. No extra phases, other than the $FeSb_2$ phase, in the pristine and solid solutions within the selected solubility range were observed. In consistence with Vegard's law, lattice parameters as well as the unit cell volume show monotonic variations with the doping levels $x$ and $y$ indicating successful substitution of Co for Fe and Te for Sb in the single-phase solid solutions. WDX reveals actual chemical compositions very close to the starting compositions of $FeSb_2$, $Fe_{1-x}Co_xSb_2$ and $FeSb_{2-y}Te_y$. The grown single crystals are of high quality as seen in the shiny surface and clearly indicated in its Laue patterns as presented in Fig. 1(b) showing the bulk-structure symmetry for a (110) plane.

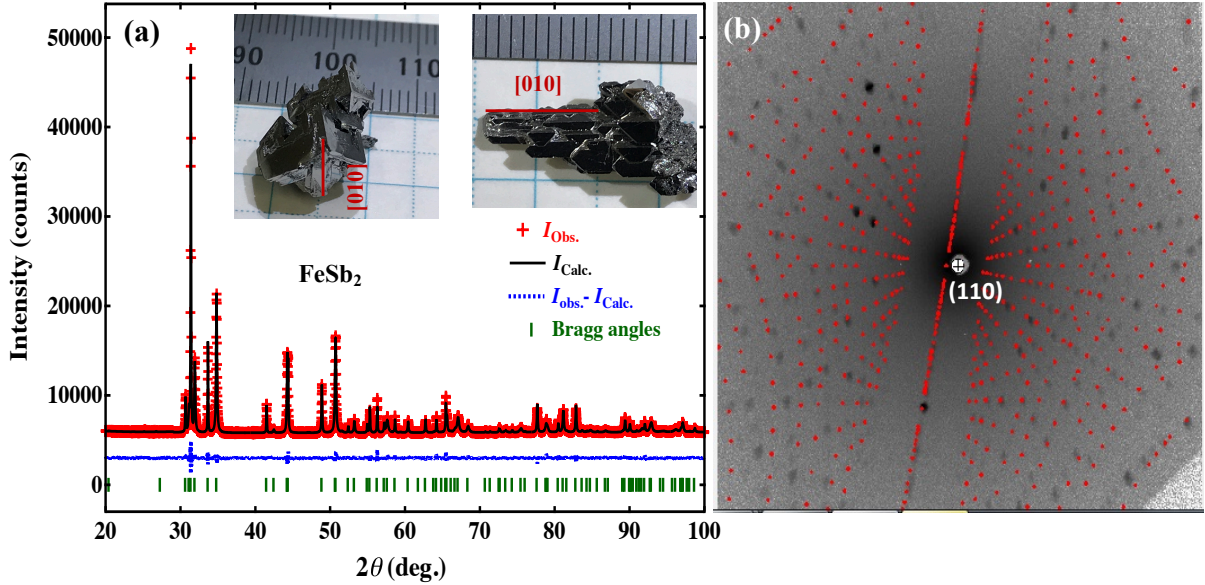

**Figure 1: (a)** The XRD pattern of powder crushed from the grown crystals of $FeSb_2$ shown with a calculated pattern by the Rietveld refinement. Insets are photos of typical crystals. **(b)** Observed and calculated Laue pattern of a (110) plane.

Figure 2(a) shows the temperature dependence of electrical resistivity, $\rho(T)$, of $FeSb_2$ measured with applying current along the three principal crystal axes. We have not observed the large anisotropy with the metal-insulator transition at 50 K along the *b*-axis reported by Petrovic et al.[4], instead and as recently reported[7], an almost isotropic semiconducting behavior down to 4 K, indicating two intrinsic energy gaps, was observed. We have estimated about 2.5 and 33 meV for the two intrinsic energy from the data below 300 K[10], which are in consistence with reported values [6,8]. The Arrhenius plot to estimate the main energy gap ($E_g$) from the linear fit in the temperature range $T$ = 70 -130 K is shown in the inset of Fig. 2(a).

Figure 2(b) shows the temperature dependences of the magnetic susceptibility, $\chi$, of $FeSb_2$ measured under applied fields $H$ parallel to the three principal crystal axes, $\chi_a$, $\chi_b$ and $\chi_c$. Similar results are reported for high quality crystals, namely anisotropic $\chi$ with almost identical behavior in *a* and *b* directions and diamagnetic to paramagnetic crossover only for $H // c$, at $T \simeq$ 70 K. Small upturns at low $T$ are attributed to possible tiny amounts of magnetic impurities. The unusual magnetic properties of $FeSb_2$ of a thermal activation type above about 50 K are similar to those of FeSi[4,13], recently discussed by using a simple semiconductor-model based on the Takahashi's theory[18,23]. As itinerant electrons in conduction bands behave like non-interacted moments due to the effect of spin fluctuations, $\chi$ of $FeSb_2$ can be assumed to come mainly from thermally excited electrons in the conduction bands and holes in the valance bands, separated by an energy gap $E_g$. The susceptibility of each electron (hole), $\chi_e$ ($\chi_h$), is proportional to $1/T$ and the number of the excited electrons (holes), $n_e$ ($n_h$), is proportional to $T^{3/2}\exp(-E_g/2k_BT)$, where $k_B$ is the Boltzmann constant. Hence, the total susceptibility $\chi(T) = 2n_e(T)\chi_e(T)$ is given as,

$$\chi(T) = AT^{1/2}e^{-E_g/2k_BT}. \quad (1)$$

Linear fitting of $\ln(\chi_{avg}T^{-1/2})$ vs. $1/T$ in the temperature range 140 K < T < 220 K shown as the inset in Fig. 2(b), where $\chi_{avg} = (\chi_a + \chi_b + \chi_c)/3$ is the polycrystalline averaged susceptibility, yields $E_g$ = 428.08 ± 13 K (36.89 ± 1.12 eV). The value is in consistence with value estimated above from $\rho(T)$ (33 meV) and agrees with reported results from both magnetization and resistivity measuremnts[6,10,17,18]. The observed magnetic and electronic transport properties indicate the high quality of our FeSb$_2$ single crystals.

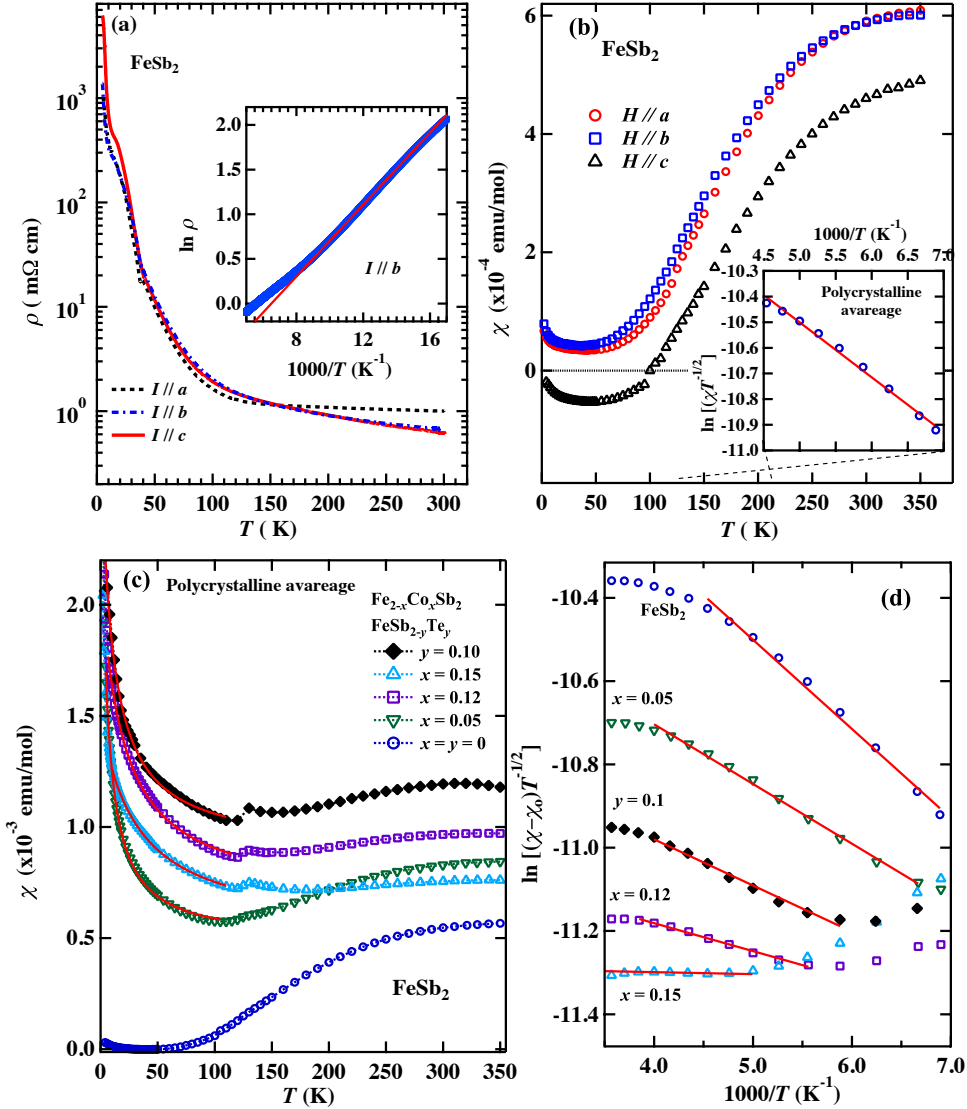

**Figure 2:** (a) The electrical resistivity $\rho(T)$ of FeSb$_2$ measured with currents applied along three principal crystal axes. Inset shows the Arrhenius plot of $\rho_b(T)$. The solid line represents a linear fit of the data. (b) Magnetic susceptibility $\chi(T)$ of FeSb$_2$ measured at $H$ = 0.1 T applied along principal crystal axes. Inset shows the plot of $\ln(\chi_{avg}T^{-1/2})$ vs. $1/T$ and its linear fit in a solid line. (c) $\chi(T)$ data of Fe$_{1-x}$Co$_x$Sb$_2$ and FeSb$_{2-y}$Te$_y$ at low doping levels. Solid lines are fits to a modified Curies-Weiss law at temperatures below 115 K. (d) $\ln(\chi_{avg}T^{-1/2}-\chi_0)$ vs. $1/T$ plots and their linear fits (solid lines) for Fe$_{1-x}$Co$_x$Sb$_2$ and FeSb$_{2-y}$Te$_y$ with $x$ and $y$ up to 0.15.

Electron doping to FeSb$_2$ influences its energy gap and results in a gradual evolution to metallic ferromagnetic state[20,22]. Figure 2(c) shows the temperature dependences of susceptibility, $\chi(T)$, of Fe$_{1-x}$Co$_x$Sb$_2$ and FeSb$_{2-y}$Te$_y$ at low doping levels below $x$ and $y$ = 0.15. The thermal activation of $\chi(T)$ gradually decreases with doping and competing paramagnetic upturns emerge below about 100 K. Anomalies observed around 130 K may be attributed to a possible magnetic transition due to this competition between the base Pauli paramagnetic and emerging magnetic states. To investigate the evolution of the energy gap by doping, from data above 140 K as in the case of FeSb$_2$, one has to consider a temperature-independent contribution, $\chi_0$, coming from the low-temperature emergent paramagnetic state. We modify equation (1) to estimate the $\chi(T)$ after low-level doping that becomes,

$$\chi(T) = \chi_0 + AT^{1/2}e^{-E_g/2k_BT} \qquad (2)$$

The temperature-independent term $\chi_0$ also contributes in the low-temperature paramagnetic state. The susceptibility behavior below about 120 K can be described by a modified Curie-Weiss (CW) law, $\chi(T) = \chi_0 + \frac{C}{T-\theta_W}$, where $C$ is the Curie constant and $\theta_W$ is the Weiss temperature. We have estimated the value of $\chi_0$ at different doping levels by fitting data below about 115 K to the modified CW law. Fittings are shown by solid lines in Fig. 2(c). Figure 2(d) shows the ln[($\chi_{avg} - \chi_0)T^{-1/2}$] vs. $1/T$ plots for Fe$_{1-x}$Co$_x$Sb$_2$ and FeSb$_{2-y}$Te$_y$ of $x$ and $y$ up to 0.15 for the high temperature data within 150 - 250 K using $\chi_0$ estimated from data below 115 K as described above. Linear fits result in a gradual decrease in the energy gap that closes at about $x$ = 0.15. Estimated values of $\chi_0$ and $E_g$ are presented in table 1.

Further electron doping to the nonmagnetic narrow-gap semiconductor results in an induced weakly itinerant electron ferromagnetism in FeSb$_2$[20,22]. Figure 3 shows the magnetic susceptibility, $M(T)/H$, and its anisotropy in Fe$_{1-x}$Co$_x$Sb$_2$ and FeSb$_{2-y}$Te$_y$ with $x$ up to 0.5 and $y$ up to 0.4 measured at a field $H$ = 0.1 T. As previously observed[20], with substituting Co for Fe (3$d$-electron doping), weak ferromagnetic states below $T_C \simeq$ 7 K with the highest magnetic moment of 0.0011 $\mu_B$ /$d$-element are observed, Fig. 3(c). However, different anisotropic behavior is observed here. The FM state emerges first in the $b$ direction at about $x \simeq$ 0.2 while in the other two directions emerges at $x$ = 0.3 and again disappears in the $a$ and $c$ directions for higher doping levels. Further, our observation of a FM state emerges in the $c$ direction at $x$ = 0.3 varies from previously reported results[20]. We attribute these differences to a sample dependence of magnetization anisotropy similar to the resistivity.

The 5p-electron doping by substituting Te for Sb results in significantly stronger ferromagnetic states than the $d$-electron doping[22]. Figure 3(d) show the temperature dependence of $M(T)/H$ of FeSb$_{2-y}$Te$_y$ with $y$ up to 0.4 at a field $H$ = 0.1 T applied along the $b$-axis. Anisotropic magnetic behavior is shown in Fig. 3(e) for $y$ = 0.4 with $b$ is the easy axis, being deferent from the $a$ easy axis in Fe$_{1-x}$Co$_x$Sb$_2$. Gradual increases of both $M$ up to 0.015 $\mu_B$/Fe, see Fig. 3(f), and $T_C$ up to 85 K are observed in contrast to the result of $d$-electron doping, with two orders of magnitude higher in both $M$ and $T_C$ for $y$ and $x \simeq$ 0.4. The inset of Fig. 3(d) shows the temperature dependences of inverse susceptibility for selected Fe$_{1-x}$Co$_x$Sb$_2$ and FeSb$_{2-y}$Te$_y$ which clarify the gradual change in the susceptibility from the thermal activated type to the CW behavior. Dashed lines are the fit to a modified CW law. Another difference is the Wiess temperature, $\theta_W \simeq$ –70 K, for $y$ = 0.4, compared to $\theta_W \simeq$ 3 K for $x$ = 0.3, which implies the coexistence of an antiferromagnetic (AF) component in $FeSb_{2-y}Te_y$. A canted AF state has been reported in this system by Hu et al[22]. Magnetic parameters of both systems, $Fe_{1-x}Co_xSb_2$ and $FeSb_{2-y}Te_y$, are presented in table 1. The observed values of $\mu_{eff}/\mu_s$ much higher than standard value of the localized system, $\mu_{eff}/\mu_s \sim 1$, indicate the itinerant-electron FM nature of the studied systems.

Figure 4 shows the magnetic phase diagram of $Fe_{1-x}Co_xSb_2$ and $FeSb_{2-y}Te_y$ ($0 < x$ and $y < 0.6$) showing the evolution of the energy gap $E_g$, closed circles for $Fe_{1-x}Co_xSb_2$ and closed square for $FeSb_{2-y}Te_y$, estimated from $M$ data. The value of $E_g$ for $FeSb_2$ is in consistence with that estimated from resistivity, open diamond from our results above and open square and triangle by Bentein et al. and Takahashi et al. respectively[6,8]. The energy gap collapses by the electron doping and closes completely at $x, y \simeq 0.15$ before the emergence of a FM metallic state that becomes strongest at around $x, y \simeq 0.5$. Suppression of the FM state towards the nonmagnetic compounds, $CoSb_2$ and $FeTe_2$, has been observed above $x = y \simeq 0.5$[20,22]. The inset shows the composition dependence of $T_C$ of $Fe_{1-x}Co_xSb_2$ and $FeSb_{2-y}Te_y$. The observed monotonic decrease of the energy gap as well as the qualitative evolution of the emergent FM with Co ($3d$ electrons) and Te ($5p$ electrons) are identical, which imply the itinerant electron character of the magnetization that depends on the electron number in the conduction band rather than the magnetically active ions (Fe and Co). A nearly ferromagnetic semiconductor is the picture proposed for $FeSb_2$ as well as FeSi in the spin-fluctuation theory[18,23]. The induced ferromagnetism in these strongly correlated low-carrier systems has been also discussed in the Kondo insulator scenario[24]. Further, the concentration dependence of the magnetization per transition metal atom in $Fe_{1-x}Co_xSb_2$, see table 1, agrees well with the observed concentration dependence in $Fe_{1-x}Co_xSi$ and accords with the prediction of a correlated band insulators model[25]. However, the magnetically ordered state and the much higher Curie temperature and spontaneous moment by $p$-electron doping in $FeSb_{2-y}Te_y$ relative to those by $d$-electron doping in $Fe_{1-x}Co_xSb_2$ system are yet to be further clarified.

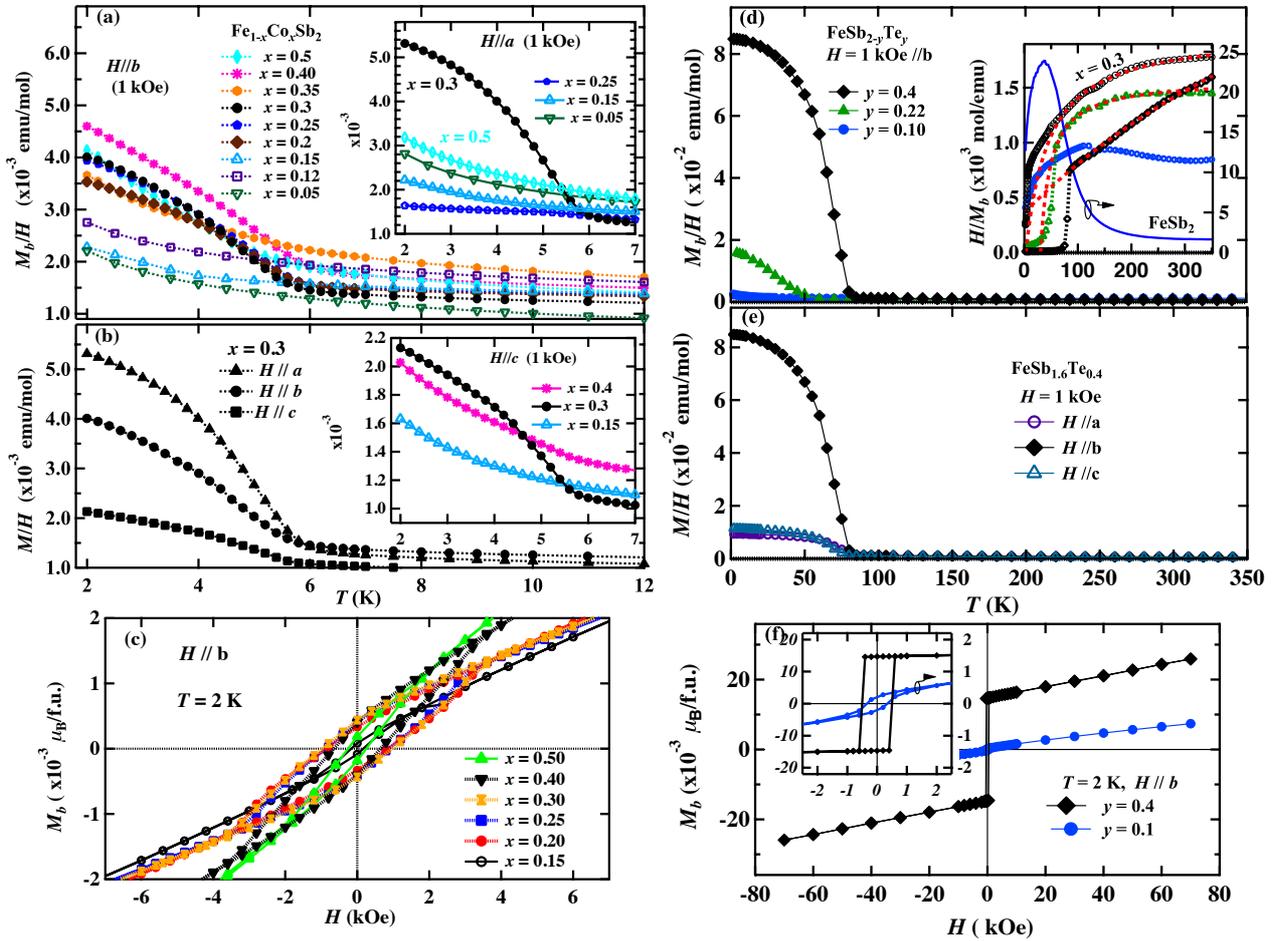

**Figure 3: (a)** Magnetic susceptibility $M(T)/H$ of $Fe_{1-x}Co_xSb_2$ with $x$ up to 0.5 under the field of 0.1 T applied along the $b$-axis. **(b)** Magnetic anisotropy of $Fe_{0.7}Co_{0.3}Sb_2$ under the field applied along the principal axes. Susceptibility data for different $x$ doping under the fields applied along $a$ and $c$ axes are shown as insets in (a) and (b), respectively. **(c)** Magnetic hysteresis in the $M$-$H$ curves of $Fe_{1-x}Co_xSb_2$ with $H // b$. **(d)** $M(T)/H$ of $FeSb_{2-y}Te_y$ with $y$ up to 0.4 under the field of 0.1 T applied along the $b$-axis. **(e)** Magnetic anisotropy of $FeSb_{1.6}Te_{0.4}$. **(f)** Magnetic hysteresis in the $M$-$H$ curves of $FeSb_{2-y}Te_y$ with $H // b$. Inset of (d) shows the inverse susceptibility of $FeSb_{2-y}Te_y$ and $Fe_{0.7}Co_{0.3}Sb_2$. Dashed lines are the fits to a modified CW law.

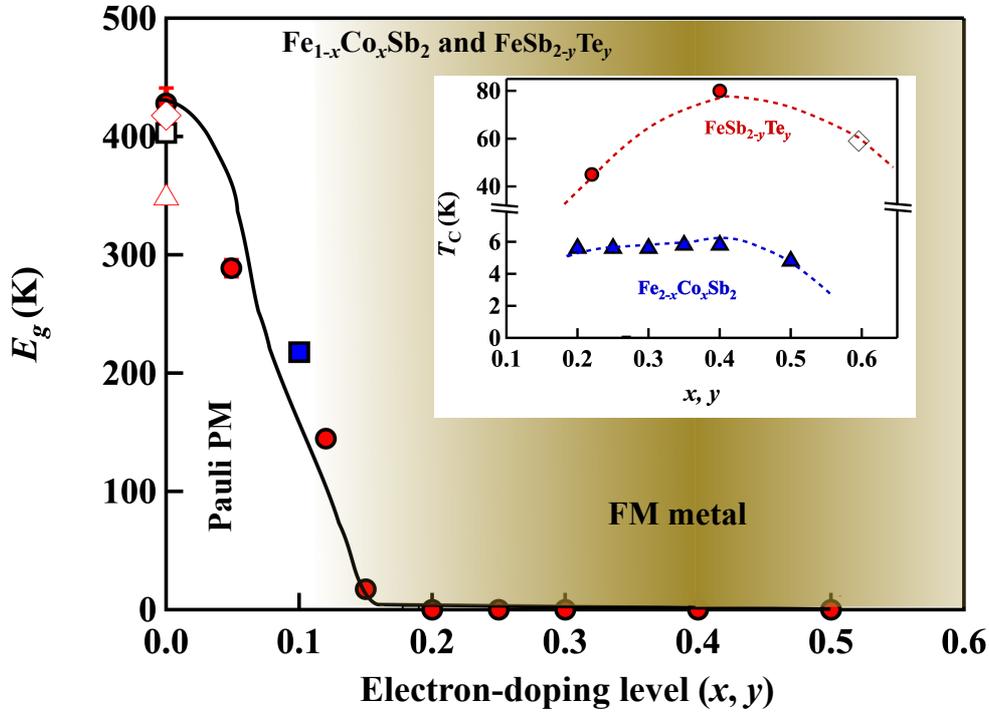

**Figure 4**: Magnetic phase diagram of $Fe_{1-x}Co_xSb_2$ and $FeSb_{2-y}Te_y$ showing the evolution of the energy gap $E_g$, closed circles for $Fe_{1-x}Co_xSb_2$ and closed square for $FeSb_{2-y}Te_y$ from magnetization data. Open diamond for $FeSb_2$ estimated from resistivity data in this work. Open square and triangle after Bentein et al. and Takahashi et al. respectively[6,8]. The inset shows the composition dependence of $T_C$ of $Fe_{1-x}Co_xSb_2$ and $FeSb_{2-y}Te_y$, $T_C$ for $y = 0.6$ (open diamond) is cited from reference[22].

**Table 1:** Magnetic parameters of FeSb$_2$ and its electron-doped solid solutions.

| x, y | $\chi_0$ (10$^{-4}$ emu/mol) | $E_g$ (K) | $T_C$ (K) | $\theta_W$ (± 3 K) | $\mu_{eff}$ ($\mu_B$/f.u.) | $\mu_s$ (10$^{-4}$ $\mu_B$/f.u.) | $\mu_{eff}/\mu_s$ |
|---|---|---|---|---|---|---|---|
| FeSb$_2$ | | | | | | | |
| 0 | 0 | 428 (13) | | | | | |
| Fe$_{1-x}$Co$_x$Sb$_2$ | | | | | | | |
| 0.05 | 4.39 | 289 (7) | | | | | |
| 0.12 | 5.65 | 145 (3) | | | | | |
| 0.15 | 5.45 | 17 (1) | | | | 0.8(2) | |
| 0.2 | 6.56 | 0 | 5.6 | 5 | 0.386(11) | 4.47(1) | 863 |
| 0.25 | 7.04 | 0 | 5.6 | 16 | 0.408(57) | 5.57(1) | 732 |
| 0.3 | 4.95 | 0 | 5.6 | 3 | 0.411(1) | 5.79(1) | 710 |
| 0.35 | 1.96 | 0 | 5.8 | 3 | 0.623(3) | 5.29(7) | 1178 |
| 0.4 | 5.56 | 0 | 5.8 | 1 | 0.602(6) | 4.94(1) | 1219 |
| 0.5 | 2.95 | 0 | 4.8 | 4 | 0.536(4) | 2.60(2) | 2062 |
| FeSb$_{2-y}$Te$_y$ | | | | | | | |
| 0.1 | 8.89 | 218 (5) | | | | 3.26(2) | |
| 0.22 | 6.35 | 0 | 45 | 30 | 0.306(1) | 17.07(1) | 179 |
| 0.4 | 2.08 | 0 | 80 | -70 | 1.180(1) | 147.42(2) | 80 |

## 4. Conclusion

We have successfully grown large single crystals of FeSb$_2$ and its electron-doped solid solutions, Fe$_{1-x}$Co$_x$Sb$_2$ (0 ≤ x ≤ 0.5) and FeSb$_{2-y}$Te$_y$ (0 ≤ y ≤ 0.4) with a simplified self-flux technique out of Sb flux. The high quality of grown single crystals has been indicated by XRD, WDX, Laue x-ray patterns, magnetization and resistivity results. The evolution from a Pauli paramagnetic state to itinerant-electron anisotropic ferromagnetic states with electron doping are observed. By employing the Takahashi's model of the itinerant-electron moment by thermal activation established in the spin fluctuation theory we have estimated the band gap, that consists with resistivity results, and its evolution with electron doping. We observed gradual decrease of the energy gap that closes with electron doping at about x = y = 0.15 followed by an emergent metallic ferromagnetic state. The magnetic phase diagram has been stablished and the results were briefly discussed in view of proposed scenarios with focus on the picture of nearly ferromagnetic narrow-gap semiconductor proposed in the spin fluctuation theory.